\documentstyle[preprint,epsf,aps]{revtex}
\begin{document}
\title{Perfect random number generator is unnecessary for 
secure quantum key distribution }
\author{Xiang-Bin Wang\thanks{email: wang$@$qci.jst.go.jp} 
\\
        Imai Quantum Computation and Information project, ERATO, Japan Sci. and Tech. Agency\\
Daini Hongo White Bldg. 201, 5-28-3, Hongo, Bunkyo, Tokyo 113-0033, Japan}

\maketitle 
\begin{abstract}
Quantum key distribution(QKD) makes it possible for two remotely separated parties
do unconditionally secure communications. In principle, the security is 
guaranteed by the uncertainty principle in quantum mechanics: if any third
party watches the key, she must disturbs the quantum bits therefore she has a risk to 
be detected. However, the security in practice is quite different,
since many of the assumptions of the ideal case do not exist. Our presently
existing secure proof of QKD protocols require the perfect random number 
generators. Actually, we can never have perfect generators in the real world.
Here we show that the imperfect random numbers can also be used for secure
QKD, if they satisfy certain explicit condition. 
\end{abstract}
Quantum key distribution(QKD) has abstracted strong interests of scientists
since it makes it possible to set up unconditional secure key between
two remote parties by principles of quantum mechanics. However, the unconditional security {\it in principle}
 does not necessarily give rise to the unconditional
security in practice, where many non-ideal factors occur.  
``The most important question in quantum cryptography is to determine how 
secure it really is''\cite{brassad}. Different from the assumed ideal
situation, there are many imperfections in realizing anything in the real 
world. Consider the case of QKD. These imperfections may include the 
channel noise, small errors in source and devices, biased random
number generators and so on. So far, the security proof with channel noise
have been given by a number of authors\cite{ekert,qkd,maye,biha,shorpre,six,BDSW,gl,chau,wang}.  The security was then
extended to the case including
small errors in source and devices\cite{ko,logo}. However, the effect of
imperfection of random numbers is still unknown.
This causes problems in practice, though it is many people's belief
that sufficiently good random numbers must also work very well.
 However, if the security is based on
such a {\it belief}, then it is still  {\it conditional
security} instead of the unconditional one and we don't know why this 
belief must be stronger than the beliefs on the assumed complexity 
of certain mathematical problems which are
the base of classical key distribution.  Therefore a strict proof is needed
here for {\it unconditional security} of QKD in the {\it real world}. 
Without an explicit analysis on the effect of imperfect random 
numbers, we don't 
know how good is sufficient therefore we have no choice but to blindly
increase the quality of random number generators. This can in principle
raise
the total cost unlimitedly. No matter how much we
have done in improving the quality of our random numbers, we still worry 
the security a little
bit, e.g., we don't know whether a bias of $10^{-10}$ or $10^{-30}$
undermines the security
severely. The best way to solve the issue is to give an explicit
study on the effect of the imperfections with certain operational
criterion. We show that, a QKD with a good imperfect random number generator
(IRNG) with certain explicit condition is secure
if the same QKD protocol with a 
perfect random number generator(PRNG) is secure.

We start from the definition of a PRNG and quantifying of an IRNG.
Consider the case that the generator produces an $\omega$-bit string,
$s$. There are $2^\omega$ possible different strings in all, we define
all $\omega$ string as set $\{s_i, i=1,2\cdots 2^\omega\}$. An
$\omega$-bit PRNG is defined
as a generator which generates every string with equal probability over
$\{s_i\}$, i.e., every string in $\{s_i\}$ have the same probability,
$2^{-\omega}$ to be generated. If a generator generates 
$\omega-$bit strings with a non-uniform probability distribution
over $\{s_i\}$, it is an imperfect random number generator,
IRNG. Specifically, we quantify the quality
of an $\omega-$bit random number generator by the value of {\it entropy}:
Suppose it generates string $s_i$ with probability $p_i$, the entropy
\begin{eqnarray}\label{epsilon}
H(s)=-\sum_{i=1}^{i=2^\omega}p_i\log_2p_i.
\end{eqnarray}   
In particular, we denote $R_{\omega,\theta}$ for an IRNG which generates
$\omega-$bit binary strings with entropy $-\sum p_i\log_2 p_i=\theta$.
In case of PRNG, the probability distribution is uniform and the above
function reaches its maximum value $\omega$. Given an IRNG, the entropy
is always less than $\omega$. Here we shall consider the case of a good
IRNG, of which the entropy value
is $\omega - \epsilon$. 
In practice, it is the case that Alice and Bob assume that they are using
a perfect random string but actually they are using an imperfect
random string which is a little bit different
from the perfect one.
It can also be the case that they only know the lower bound
of the quality of their generator, say $H(s)\ge \omega - \epsilon$,
but they don't know the explicit pattern of the strings
and they have no way to change the generator to a perfect one.
 However, we should assume the worst case that Eavesdropper (Eve) knows
the pattern of their string though they themselves don't know it.
For example, a very smart Eve could find out the pattern from the history
of the data. Also, in QKD, we assume Eve knows the protocol itself. This
means Eve knows the specific status of all the devices involved, including
the random number generators. Therefore, given the existing
security proofs with PRNGs, 
we still worry a little bit that Eve could take advantage
of her knowledge about the IRNG being used in the protocol and obtain
a larger amount of information than the theoretical upper bound in the case
PRNGs are used.
Our purpose is to see whether Eve
can obtain significantly large information to the final key generated
by certain QKD protocol with good IRNG, if Eve's information to the final
key by the same protocol with PRNG is in principle
bounded by a very small value.
 In all existing QKD protocols, both Alice and Bob needs some
random numbers for the task. For example, in BB84 protocol\cite{BB}, Alice needs the random numbers 
to prepare the initial quantum state $|0\rangle$ or $|1\rangle$
for each qubits; he needs
 to choose a subset of the qubits for the error test and he also needs
random numbers  to
make error correction, privacy amplification finally. 
For all these issues,
Alice only needs to prepare an $\omega_a-$bit binary random string $s$ in
the beginning, if the protocol needs 
$\omega_a$ random bits in all at Alice's side.
In carrying out the protocol, Alice just reads $s$ from left to right, 
whenever a random bit is needed.
Bob also needs random
numbers to determine his measurement bases $(\{|0\rangle,|1\rangle\}$
or $\{\pm=\frac{1}{\sqrt 2}(|0\rangle\pm |1\rangle)\})$. 
We at this moment assume
Bob has perfect random numbers while Alice does not. After we complete the proof of our {\bf Theorem}, 
we extend it to the case that Bob does not have
perfect random numbers either.
 We shall first show the following theorem:\\
{\it {\bf Theorem:}  Given any QKD protocol $P$, suppose Bob always uses
a PRNG and
Eve knows what random generators are used by Alice and Bob,
if Eve's information is bounded by $\epsilon_0$ in the case that
 Alice uses a PRNG, then Eve's information to the $k-$bit final key
is bounded
by $$\epsilon_0+(4k+1)
\sqrt {\epsilon_A/2}+{\rm {\bf O}}(\epsilon_A^{3/2})+
{\rm {\bf O}}(\epsilon_A)$$ 
to the  final key
in the case Alice uses an IRNG, $R_{\omega_a,\epsilon_A}$.
}\\ 
 For clarity let us first recall the 
theorem of Holevo bound\cite{hole,nil}.\\
 Clare announces the following facts:
He will Alice an $\omega_a-$bit state which can be
either $\rho_0$ or $\rho_1$, with equal probability.
He sets his bit value $X=0$ if he passes $\rho_0$ to Alice, and
$X=1$ if he passes $\rho_1$ to Bob. It's known to all parties that
$\rho_0 =(2^{-\omega_a})\sum_{i=1}^{2^{\omega_a}}|s_i\rangle\langle
s_i|$
 and   
$\rho_1 =\sum_{i=1}^{2^{\omega_a}}p_i |s_i\rangle\langle s_i|$. 
State $|s_i\rangle$
is a product state of $\omega_a$ qubits with each of them being prepared
in $\{|0\rangle,|1\rangle$ basis. String $s_i$ gives the full information
of state of each qubits, e.g., if $s_j=01010011\cdots 10$, then 
$|s_j\rangle=|01010011\cdots 10\rangle$. In such a case, using Holevo's 
theorem\cite{hole} we 
find that Alice's information to bit $X$ is bounded by
\begin{eqnarray}\label{hb}
h=H(\bar p)-\frac{\omega_a}{2}-\frac{1}{2}H(p)
\end{eqnarray}
and $p={p_i}$
In fact, in this case, if Alice directly observes each qubits 
in $\{|0\rangle,|1\rangle\}$
basis she can reach the upper bound of information to $X$. 
With a little bit calculation, one immediately obtain the fact that
\begin{eqnarray}\label{bh}
h \le  \epsilon_a/2.
\end{eqnarray}
In what follows we shall show that, with the restriction by Holevo's theorem,
Eve's information to the final key must be negligible in a QKD with good
IRNG $R_{\omega_a,\epsilon_A}$, if her information is in principle
negligible in the same QKD protocol
with perfect random numbers.
We now consider the following game\\
{\bf Game G:} 
Clare announces that he will pass Alice an $\omega_a-$bit state which can be
either $\rho_0$ or $\rho_1$, with equal probability, and
he sets his bit value $X=0$ if he passes $\rho_0 =(2^{-\omega_a})\sum_{i=1}^{2^{\omega_a}}|s_i\rangle\langle
s_i|$ to Alice, and
$X=1$ if he passes state 
$\rho_1 =\sum_{i=1}^{2^{\omega_a}}p_i |s_i\rangle\langle s_i|$. 
 to Bob.   
Alice measures each qubits of the state from Clare in $\{|0\rangle,|1\rangle\}$ basis
and obtain a classical string $s$. Using $s$ as the random string Alice runs QKD protocol
$P$ with Bob.
At this moment Eve attacks the protocol just as if she were a real eavesdropper.  
If the protocol does not pass the error test, Alice gives it up and uses
string $s$ to obtain the information about $X$, in such a case she can reach 
the Holevo bound. If the protocol passes the error test, they continue and 
set up a $k-$bit final key $Y$ and then Alice announces  it. 
In such a case,
Eve can obtain information about $X$ by reading the final key, 
and Eve reports her information 
about $X$ to Alice and Alice uses this as her own information about $X$.
Obviously, Eve's information about $X$ must also be bounded by $h$, otherwise
the result of our game violates Holevo's theorem.
 
 Suppose scheme $T$ is the optimal
attack to QKD protocol $P$ with imperfect random string whose Shannon entropy
is $\epsilon_A$ (but $T$ is not necessarily optimal to the same 
protocol with {perfect}
random string).  Suppose Eve attacks $Y$ by scheme $T$.
Without any loss of generality, $T$ has the following property:
if string $s$ used by Alice is perfectly random, Eve acquires information
$\epsilon'$ about $Y$. 
If $s$ is 
from
generator $R_{\omega_a,\epsilon_A}$, Eve's information about the final key
is optimized, we denote it by $\eta$ in such a case. In our game 
$R_{\omega_a,\epsilon_A}$ corresponds to the case $X=1$. 
Intuitively, $\eta$ should not be too large given $\epsilon'$
being very small, since otherwise after Alice 
announces $Y$, Eve may easily see whether her actual information about $Y$
prior to the announcement is $\eta$ or $\epsilon'$ therefore she can
access
an unreasonably large amount of 
information about Clare's bit $X$. 
After attack $T$, Eve has 2 sets of probability distribution
$P=\{P_i\},Q=\{Q_i\}$ about the $k-$bit final key $Y$, 
conditional on $X=0,1$, respectively. Before reading
final key $Y$, these two sets of distribution about $Y$ have equal probability. More specifically, after reading the final key $Y$, the two distributions 
$P$ and $Q$ can be different. Therefore probability of $X=0$ and
$X=1$ can also be different after Eve reads $Y$. That is to say, in reading
$Y$, Eve may obtain different probabilities for 
the probability distribution $P$ and $Q$.\\
For simplicity, the two probabilities for a specific possible
final key, $Y_j$. 
Consider one possible way for Alice to violate the Holevo's theorem:
If the final key is not $Y_j$, she disregard the QKD result just
uses string $s$ itself and obtain 
 information about $X$ in the amount of Holevo bound. If the final key
is $Y_j$, she announces $Y_j$ and uses Eve's information about $X$ as
her own information. Therefore,  Eve's two probabilities ($P_j,Q_j$)
about any $Y_j$
cannot be too different, otherwise Alice has non-zero chance to violate
Holevo's theorem. Specifically,
we have the following restriction
\begin{eqnarray}\label{yj}
I_E(X:Y_j)=1-\frac{1}{2}H(P_j')-\frac{1}{2}H(Q_j')
\le h.
\end{eqnarray}
Here $P_j'=\frac{P_j}{P_j+Q_j}$, $Q_j'=\frac{Q_j}{P_j+Q_j}$
and $H(t)=-t\log_2 t-(1-t)\log_2 t$. For simplicity
we shall use $\log$ instead of $\log_2$ hereafter.
The above formula is equivalent to
\begin{eqnarray}
I_E(X:Y_j)=1+\frac{1}{2+\delta}\log\frac{1}{2+\delta}+
\frac{1+\delta}{2+\delta}\log\frac{1+\delta}{2+\delta}\le h.
\end{eqnarray} 
Here $\delta_j$ is defined by $\delta_j=Q_j/P_j-1=Q_j'/P_j'-1$.
After a further reduction we obtain
\begin{eqnarray}
I_E(X:Y_j)=-\frac{1}{(2+\delta_j)}
\log(1+\delta_j/2)+\frac{1+\delta_j}{2+\delta_j}
\log\left(1+\frac{\delta_j}{2+\delta_j}\right)
\end{eqnarray}
If $\delta_j\ge 0$, we have
\begin{eqnarray}
I_E(X:Y_j)\ge -\frac{\delta_j}{2(2+\delta_j)}
+\frac{(1+\delta_j)}{(2+\delta_j)}\left[\frac{\delta_j}{2+\delta_j}
-\frac{\delta_j^2}{2(2+\delta_j)^2}\right]\ge 
\frac{\delta_j ^2}{2(2+\delta_j)^3};
\end{eqnarray}
if $\delta_j< 0$, we have
\begin{eqnarray}
I_E(X:Y_j)\ge -\frac{1}{(2+\delta_j)}(\delta_j/2+\delta_j^2/8)
+\frac{\delta_j(1+\delta_j)}{(2+\delta_j)^2}
\ge \frac{2\delta_j^2-\delta_j^3}{8(2+\delta_j)^2}\ge  
\frac{2\delta_j^2}{8(2+\delta_j)^2}.
\end{eqnarray}
In any case, we have
\begin{eqnarray}
|\delta_j| \le |\Delta =4[1+{\rm {\bf O}}(\sqrt h)]\sqrt h
\end{eqnarray}
for any $j$, given the restriction of formula (\ref{yj}). 
With this formula, we can now
calculate the lower bound of $H(Q)$, the entropy to the whole $k-$bit
final key, given distribution $Q=\{Q=i\}$.
\begin{eqnarray}
H(Q)=H(\{P_i(1+\delta_i\})=-\sum_{i}P_i(1+\delta_i)\log[P_i(1+\delta_i)]
\nonumber\\
\ge (1-|\Delta|)H(P)- (1+|\Delta|)\log (1+|\Delta|)\nonumber\\
\ge k-\epsilon' -
(4k+1) [1+{\rm {\bf O}}(\sqrt h)]\sqrt h -  {\rm {\bf O}}(h). 
\end{eqnarray} 
Since $h\le \epsilon_A/2$, we have
 \begin{eqnarray}
\eta
\le \epsilon' +
(4k+1) [1+{\rm {\bf O}}(\sqrt {h})]\sqrt {h} -  {\rm {\bf O}}(h). 
\end{eqnarray} 
Note that we always have $\epsilon'\le \epsilon_0$, since $\epsilon_0$ is the upper bound
of $any$ attacks to a QKD protocol with PRNG, $\epsilon'$ is the information through $T$,
which is an optimized attack to QKD with IRNG, but not necessarily also an optimized attack
to the same QKD protocol with PRNG.   Therefore we complete the proof
of our {\bf theorem} by
replacing  $\epsilon'$
with $\epsilon_0$. 
Now we consider the case that Bob's random string is also
imperfect, say, he uses an IRNG $R_{\omega_b,\epsilon_B}$. 
Since we have already known that Eve's information is bounded
by  
$ \epsilon_0 +
(4k+1) [1+{\rm {\bf O}}(\sqrt {h})]\sqrt {h} -  {\rm {\bf O}}(h)$ we
the case that Alice uses IRNG $and$ Bob uses PRNG,
we now just consider game $G'$ where
David  passes Bob an $\omega_b-$qubit state
$|\rho_0'\rangle=\sum_{i=0}^{2^{\omega_b}}|s_i\rangle\langle s_i|$
if he sets $X=0$
and passes Bob an $\omega_b-$qubit state 
$|\rho_1'\rangle=\sum_{i=0}^{2^{\omega_b}}|s_i\rangle\langle s_i|$
if he sets $X=0$
and passes Bob an $\omega_b-$qubit state 
$|\rho_1'\rangle=\sum_{i=0}^{2^{\omega_b}}p_i'|s_i\rangle\langle s_i|$
if he sets $X=1$. Similarly to the proof for our theorem, 
we have the following corollary: Suppose Eve always knows what random number generators are used
by Alice and Bob in  a certain QKD protocol $P$. If Eve's information is upper
bounded by $\epsilon_0$ to the  final key
in the case PRNGs are used, then Eve's information is 
upper bounded by 
$\eta \le \epsilon_0   +
(4k+1)\sqrt { \epsilon_A/2} 
 +(4k+1)\sqrt {\epsilon_B/2} +{\rm{\bf O}}(\epsilon_A^{3/2})
+{\rm{\bf O}}(\epsilon_A)
$ to the  $k-bit$ final 
key in the case Alice uses IRNG $R_{\omega_a, \epsilon_A}$ and Bob uses 
IRNG $R_{\omega_b,\epsilon_B}$. \\
We conclude that, {\it
if a QKD protocol is secure with perfect random numbers 
being used, it must be 
also secure with exponentially small imperfections in the random numbers.
Therefore perfect random nember generators wchich never exist
in the real world  are not necessary for secure QKD.}
\\{\bf Acknowledgement:} I thank Prof Imai H for support. 
I also thank K. Matsumoto, M. Hayashi and M. Hamada for discussions.


\begin{thebibliography}{99}
\bibitem{brassad} G. Brassard and  C. Crepeau, SIGACT News 27(No. 3), 13(1996).
\bibitem{ekert}C. H. Bennett, G. Brassard, S. Popescu, B. Schmacher, J. Smolin, and
W. K. Wooters, Phys. Rev. Lett., 76, 722(1996);
 D.~Deutsch, A.~Ekert, R.~Jozsa, C.~Macchiavello,
S.~Popescu, and A.~Sanpera, Phys.~Rev.~Lett., 77, 2818(1996);
  Erratum
Phys.~Rev.~Lett. {\bf 80}, 2022 (1998).
\bibitem{qkd} H.-K.~Lo and H.~F.~Chau,  Science,
283, 2050(1999).
\bibitem{BB}
C. H. Bennett and G. Brassard, 
{\em Proceedings of IEEE International Conference on Computers, 
Systems and Signal Processing, Bangalore, India, 1984},  (IEEE Press,
1984), pp. 175--179;
C.H. Bennett and G. Brassard,
IBM Technical Disclosure Bulletin {\bf 28}, 3153--3163 (1985).
\bibitem{maye} D. Mayers,
                 J. Assoc. Comput. Mach. {\bf 48}, 351 (2001).
\bibitem{biha} E. Biham, M. Boyer, P.O. Boykin, T. Mor, and
               V. Roychowdhury, in {\it Proceedings of the
               Thirty-Second Annual ACM Symposium on Theory
               of Computing} (ACM Press, New York, 2000),
               pp.715-724, quant-ph/9912053.
\bibitem{shorpre} P. W. Shor and J. Preskill, Phys. Rev. Lett., vol. 85,441(2000). 
\bibitem{six}H.-K. Lo, quant-ph/0102138.
\bibitem{BDSW} C. H. Bennett, D. P. DiVincenzo, J. A. Smolin,
and W. K. Wootters, Phys. Rev. A54, 3824(1996).
\bibitem{gl} D. Gottesman and H.-K. Lo, IEEE Transactions on
 Information Theory, 49, 457(2003).
 \bibitem{chau} H. F. Chau, Phys. Rev. A66, 060302(R) (2002).
\bibitem{wang} X. B. Wang, Phys. Rev. Lett., 92, 077902(2004).
\bibitem{ko}M, Koashi and J. Preskill, Phys. Rev. Lett. 90, 057902(2003)
\bibitem{logo}D. Gottesman, H.-K. Lo, N. Lutkenhaus, and J. Preskill, quant-ph/0212066. 
\bibitem{hole} Statiscal Capacity of a quantum communications channel,
Problems of Inf. Transm, 5(4):247-253(1979).
\bibitem{nil}M. A. Nielsen and I. L. {\it Chuang, Quantum Computation and Quantum Information}, Cambridge University Press, 2000.
\end{thebibliography}
\end{document}